\documentclass[aps,prb,twocolumn,superscriptaddress]{revtex4-1}
\usepackage{times}
\usepackage{graphicx}
\usepackage[caption=false]{subfig}
\usepackage{dcolumn}
\usepackage{bm}
\usepackage{color}
\usepackage{amsmath}
\usepackage{amssymb}
\usepackage{braket}

\usepackage{graphicx}
\usepackage{epstopdf}
\usepackage{dcolumn}
\usepackage{color}
\usepackage{amsmath,amssymb}
\usepackage[english]{babel}
\usepackage{bm}

\usepackage[hyperindex]{hyperref}
\hypersetup{colorlinks,citecolor=blue, filecolor=blue,linkcolor=blue,urlcolor=blue}

\graphicspath{{./figures/}}
\usepackage{multirow}
\usepackage{rotating}
\usepackage{array}

\newcommand{\bpm}{\begin{pmatrix}}
\newcommand{\epm}{\end{pmatrix}}

\newcommand{\be}{\begin{equation}}
\newcommand{\ee}{\end{equation}}
\newcommand{\beq}{\begin{eqnarray}}
\newcommand{\eeq}{\end{eqnarray}}

\DeclareMathOperator{\im}{Im}

\DeclareMathOperator{\tr}{tr}

\newsavebox{\mybox}

\begin{document}

\title{Interface states at the boundary between ABC and ABA multilayer graphene structures}

\author{Maxence Grandandam}
\affiliation{Institut de Physique Th\'eorique, Universit\'e Paris Saclay, CEA
CNRS, Orme des Merisiers, 91190 Gif-sur-Yvette Cedex, France}
\author{Sarah Pinon}
\affiliation{Institut de Physique Th\'eorique, Universit\'e Paris Saclay, CEA
CNRS, Orme des Merisiers, 91190 Gif-sur-Yvette Cedex, France}
\author{Cristina Bena}
\affiliation{Institut de Physique Th\'eorique, Universit\'e Paris Saclay, CEA
CNRS, Orme des Merisiers, 91190 Gif-sur-Yvette Cedex, France}
\author{Catherine Pepin}
\affiliation{Institut de Physique Th\'eorique, Universit\'e Paris Saclay, CEA
CNRS, Orme des Merisiers, 91190 Gif-sur-Yvette Cedex, France}

\begin{abstract}
We study the interfaces between ABC and ABA regions in multilayer graphene, in particular we consider regions in which the transition between the ABA and ABC structure arises due to the local compression in one of the graphene layers for a zigzag interface, and to a local shear of the atoms for the armchair interface. When we consider an infinite ribbon configuration we find that interface bands form in these regions, more pronouncedly for the armchair-type interfaces. We note that these states are not topological. For fully-finite-size structures the continuous interface bands transform into sets of quantized levels.

\end{abstract}

\maketitle

\section{Introduction}

In the recent years there has been a lot of interest in the formation of edge states and interface states in graphene. The recent realization of different kinds of twisted graphene stacks also brought into focus the important differences between regions with different stacking\cite{Cao2018a,Cao2018b,Shen2020,Liu2020,Yankowitz2019}. Traditionally graphene has been shown to exhibit non-topological edge states along the zigzag and bearded edges, for both monolayer and multilayer configurations \cite{Geim2007, CastroNeto2009, Castro2008}. Moreover, topological states have been predicted to arise at the interface between AB and BA bilayer graphene regions\cite{Martin2008,Yin2016,Vaezi2013,Zhang2013}, as well as in samples obtained by twisting the two graphene layers by a certain angle\cite{Kerelsky2019,Huang2018} or in rhombohedral-stacked graphene films\cite{Slizovskiy2019,Xiao2011}. Topological edge states have also been observed in trilayer graphene ribbons subject to a transverse electric field\cite{Jung2011,Li2012}. However no study of the interfaces between ABA and ABC graphene regions has been performed.

\begin{figure}[ht]
\centering
\subfloat{\includegraphics[width=0.7\columnwidth, trim=0 40 0 50, clip]{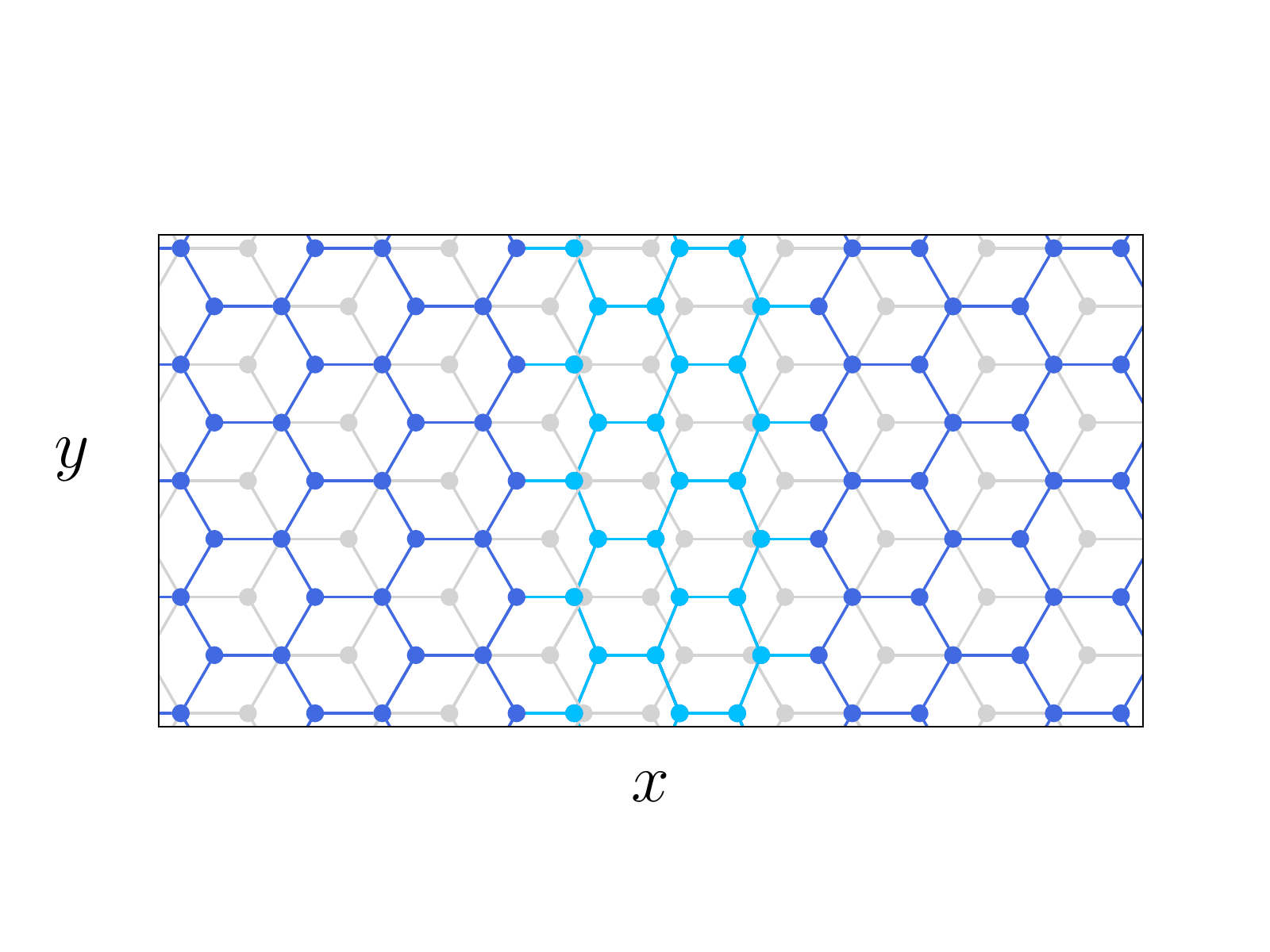}}
\\
\subfloat{\includegraphics[width=0.73\columnwidth, trim=0 0 0 50, clip]{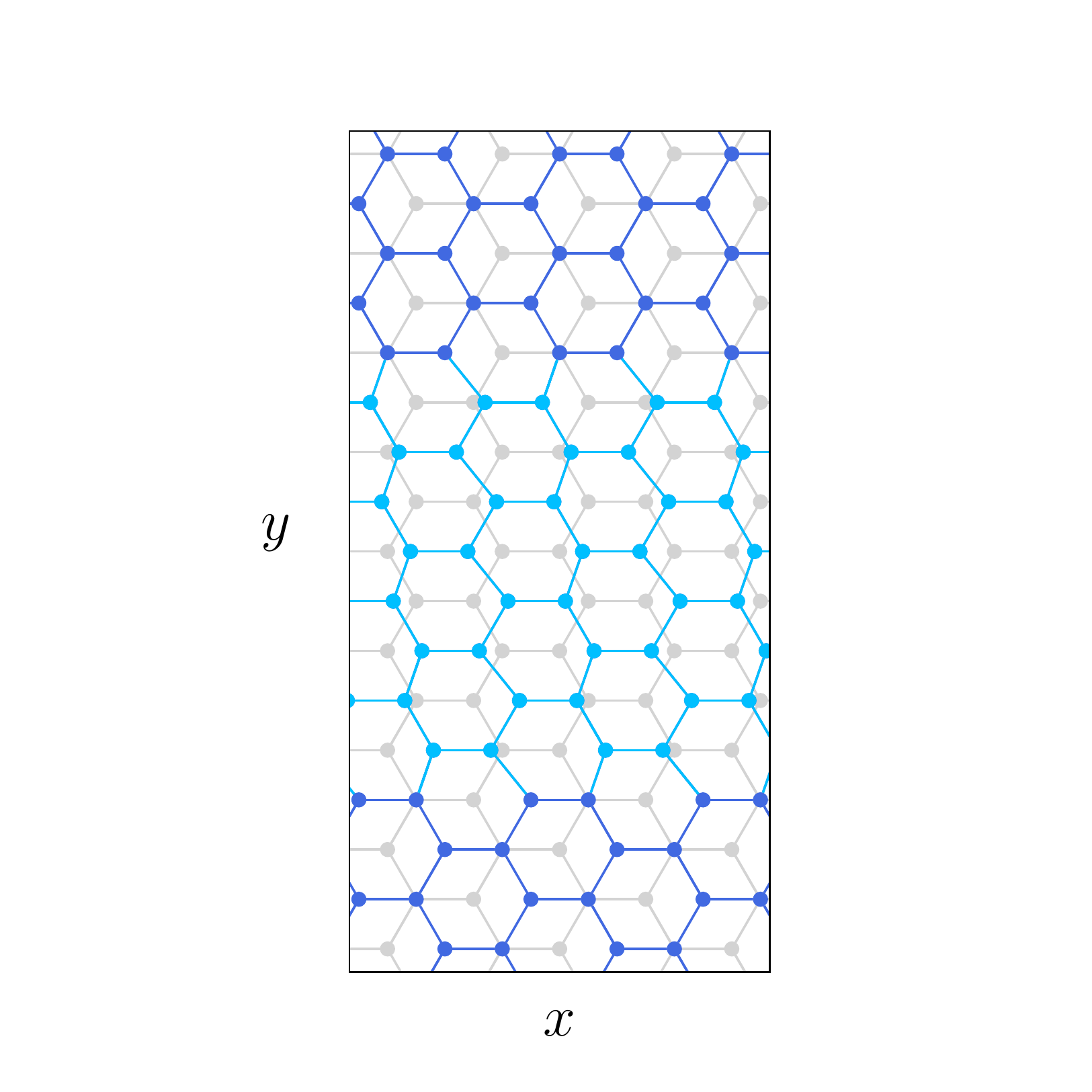}}
\caption{\textbf{(a)} A local compression in the top graphene layer for a zigzag configuration. Inside the interface region (light blue), the distance between atoms is shortened along the $x$ direction, such that on the right side of the interface all atom positions have been shifted by $-a_0 \hat{x}$ compared to those on the left side. As we can see, the stacking of the top layer relative to the underlying layer (in gray) is different on the two sides of the interface yielding a transition between ABA and ABC stacking in the trilayer system. \textbf{(b)}  A local shear of the top graphene layer atoms for an armchair configuration. Inside the interface region the $x$ coordinate of the atoms is shifted, such that on the upper side of the interface all atom positions have been shifted by $-a_0 \hat{x}$ compared to those on the lower side, yielding once more a ABA/ABC interface. We will take the impurity to be $5$ unit cells wide in the remaining of the paper.}
\label{fig:fault}
\end{figure}

In this work we study such interfaces, in both infinite ribbon configuration and finite-size configuration, for both zigzag and armchair interfaces. Inspired by the paper of Delavignette and Amelinckx \cite{Delavignette1962}, we model the transition between the ABA and ABC regions via a local compression in one of the graphene layers for a zigzag interface, and via a local shear of the atoms for the armchair interface, which generates a region of few atom-width which constitues the interface region (see Fig.~\ref{fig:fault}). The two stacking orders are known to exhibit very different physical properties\cite{Koshino2010,Serbyn2013,Guinea2006,Lin2017,Zhang2010,Min2008}, and the impact of the geometry of the edge on the properties of graphene stacks has also been studied previously\cite{Li2010}. We use a numerical tight-binding method to study the formation of states inside the stacking fault, and we find that indeed, in the infinite ribbon configuration, discrete bands of dispersing states localized inside interface regions form both in the case of zigzag and armchair edges. To test this result, we plot the ribbon band structure for various system widths, the weight of the bands as a function of position, as well as the local density of states as a function of energy and position. We also find that, when considering finite-size ribbons, the interface bands become quantized and give rise to discrete edge energy levels localized on the defect.

Our paper is organized as follows: in Section II we present a review of the band structure and the density of states (DOS) for infinite bulk ABC and ABA graphene trilayers, in Section III we present the formation of interface bands in infinite ribbons with a zigzag interface between ABA and ABC regions, in Section IV we repeat the analysis for an armchair interface, in Section V we consider a fully-finite-size system, and we conclude in Section VI.

\section{Band structure and local density of states for bulk ABA and ABC graphene}
We begin by reviewing the band structure and the local density of states for infinite ABA and ABC trilayer graphene systems. The tight-binding Hamiltonians for these honeycomb lattices are given by:

\begin{align}
\mathcal{H}(\textbf{k}) &= t_0 \sum_{l = 0,1,2} t(\textbf{k}) c_{\textbf{k}, l, A}^\dagger c_{\textbf{k}, l, B}^{} + h_{inter} + \text{h.c.},
\label{eq:hamiltonianBernal}
\end{align}
with
\begin{equation}
t(\textbf{k}) = 1 + 2 e^{- i \frac{3}{2} a_0 k_x} \cos \left(\frac{\sqrt{3}}{2} a_0 k_y \right) ,
\end{equation}
\begin{equation}
h_{inter} = t_1 \left( c_{\textbf{k}, 0, B}^\dagger c_{\textbf{k}, 1, A}^{} + c_{\textbf{k}, 1, A}^\dagger c_{\textbf{k}, 2, B}^{} \right)
\end{equation}
for ABA stacking and
\begin{equation}
h_{inter} = t_1 \left( c_{\textbf{k}, 0, B}^\dagger c_{\textbf{k}, 1, A}^{} + c_{\textbf{k}, 1, B}^\dagger c_{\textbf{k}, 2, A}^{} \right)
\end{equation}
for ABC stacking. The operator $c_{\textbf{k}, l, \alpha}^\dagger$ creates an electron on layer $l$ and sublattice $\alpha$, and $a_0$ is the distance between two neighboring atoms in a honeycomb lattice. In what follows, we take the two hopping parameters to be $t_0 = 2.6$, $t_1 = 0.34$.

\begin{figure}[h]
\centering
\subfloat{\includegraphics[width=0.49\linewidth, trim=0 0 0 10, clip]{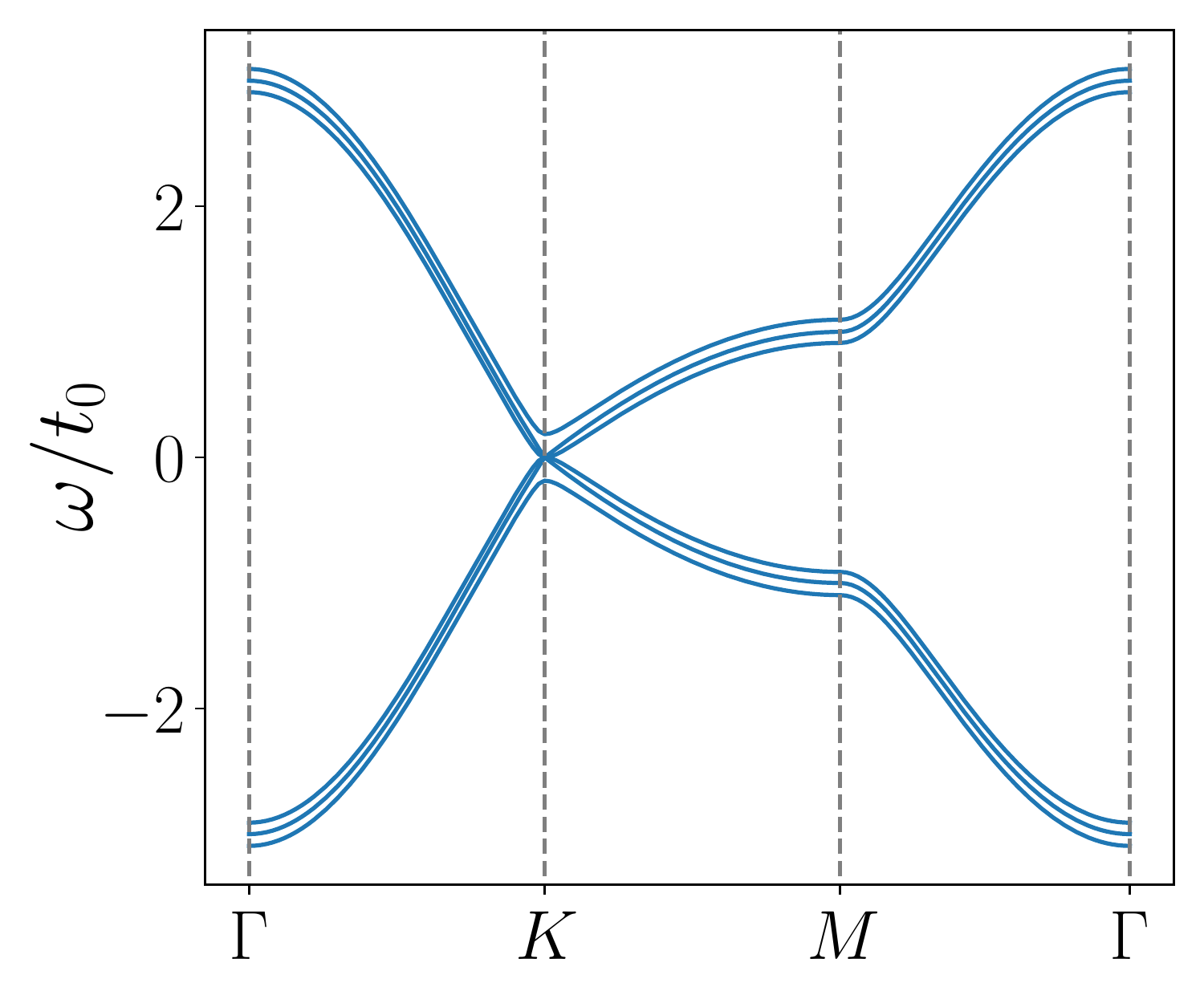}}
\subfloat{\includegraphics[width=0.49\linewidth, trim=0 0 0 10, clip]{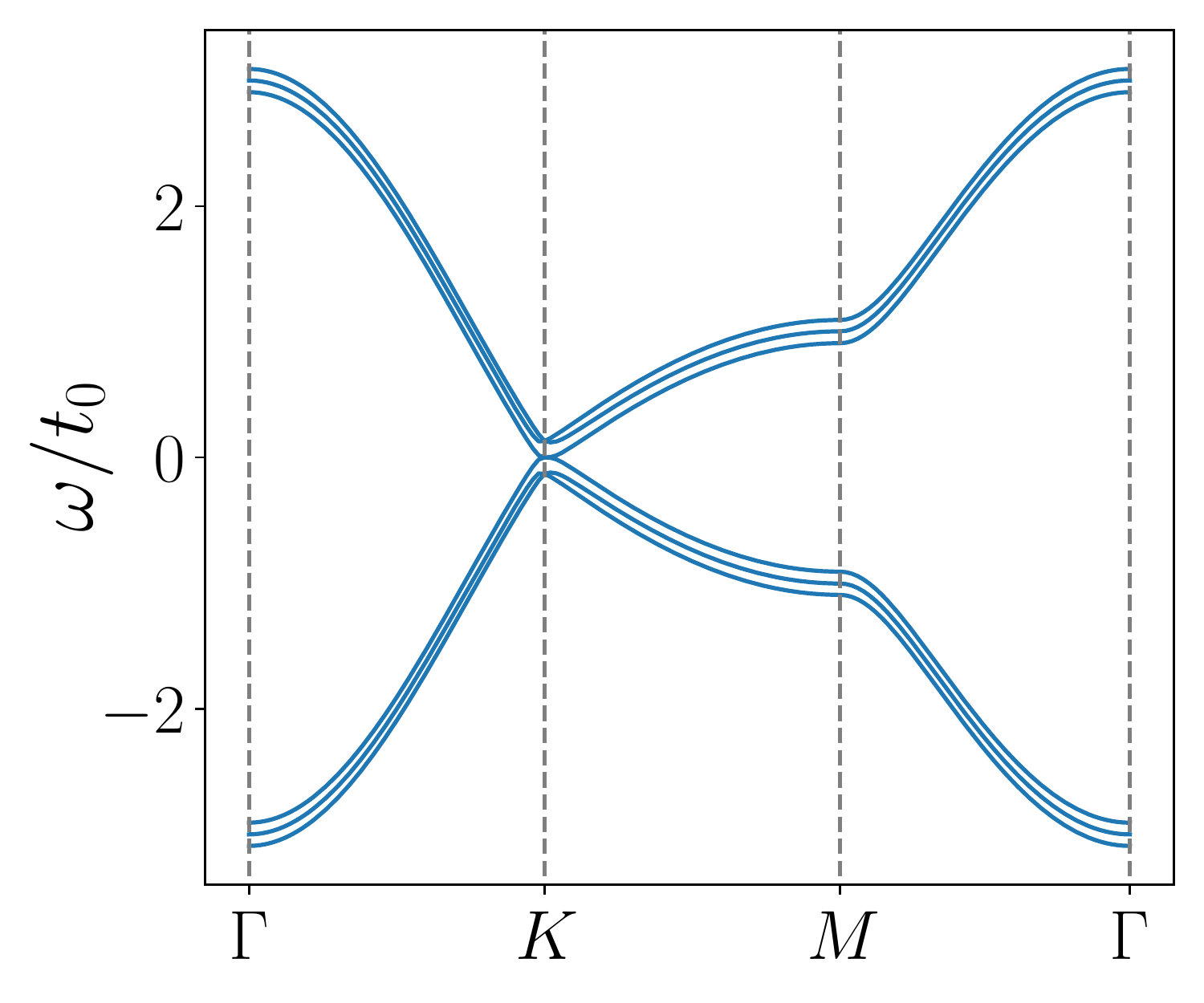}}
\qquad
\subfloat{\includegraphics[width=0.49\linewidth]{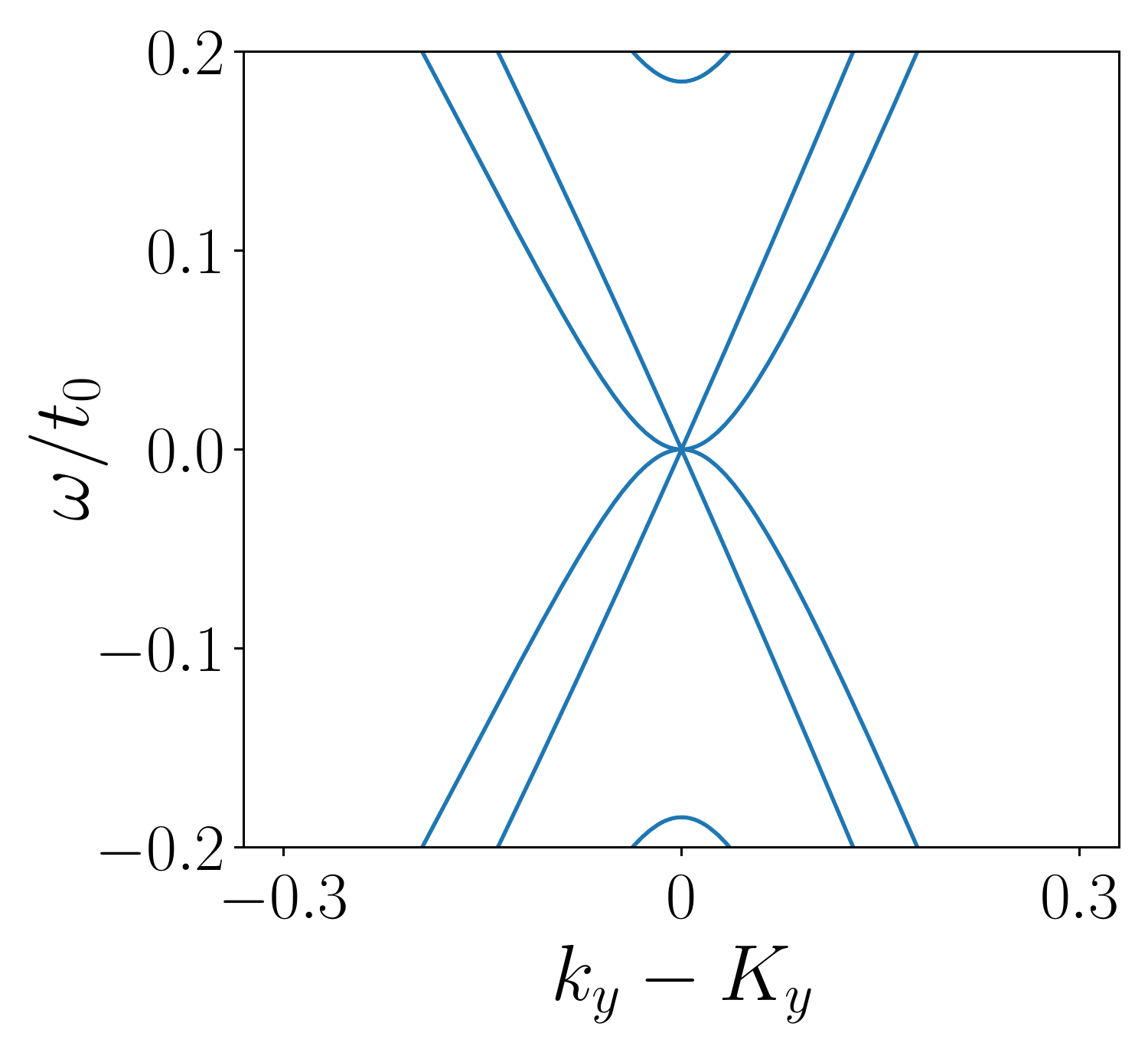}}
\subfloat{\includegraphics[width=0.49\linewidth]{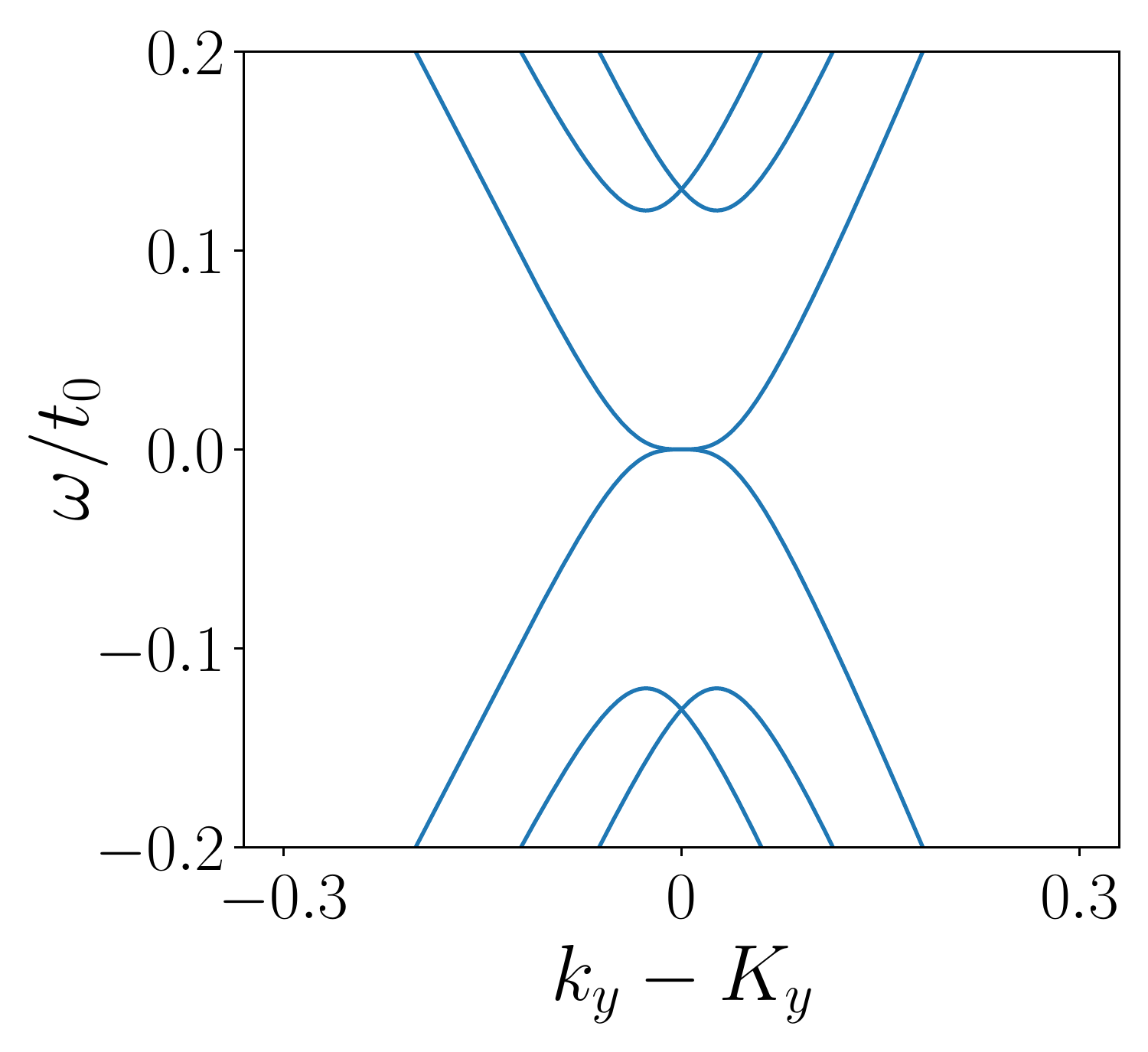}}
\caption{Bulk energy bands for ABA (left) and ABC (right) trilayer graphene. The lower plots are zooms around the Dirac point.}
\label{fig:bulkbands}
\end{figure}

The band structures obtained by diagonalizing these Hamiltonians are presented in Fig.~\ref{fig:bulkbands}. We note that the band structure for the ABA trilayer graphene contains one linear band and one quadratic band, while the ABC band structure contains two quasi-quadratic bands with an energy gap of about $0.25 t_0$ as well as a lower band which is not entirely quadratic but flattens close to the Dirac point (energy dispersion of $k^3$) creating the precursor of a flat band. This plays an important role in differentiating ABA and ABC stacking in multilayer systems, as the general dispersion for ABC stacking is of the order $k^N$ with $N$ the number of layers. This leads to an enhanced density of states at low energy compared to the ABA case (due to the ``flat bands'');  the larger the number of layers, the larger the flat region and its importance \cite{Pierucci2015, Yacoby2011}.

We define the Matsubara Green's function (GF) as: $G(\textbf{k}, i \omega_n) = \left[i \omega_n - \mathcal{H}(\textbf{k})\right]^{-1}$, where $\omega_n$ denote the Matsubara frequencies. The retarded GF $\mathcal{G}(\textbf{k}, \omega)$ is then obtained by the analytical continuation of the Matsubara GF by setting $i \omega_n \rightarrow \omega + i\delta$ with $\delta \rightarrow 0^+$. By integrating $\mathcal{G}(\textbf{k}, \omega)$ over the Brillouin zone, we can obtain the DOS as a function of energy: 

\begin{equation}
DOS(\omega) = -\frac{1}{\pi} \int \frac{d\textbf{k}}{(2\pi)^2} \im \tr \mathcal{G}(\textbf{k}, \omega).
\end{equation}

\begin{figure}[h]
\centering
\subfloat{\includegraphics[width=0.49\linewidth, trim=10 0 0 0, clip]{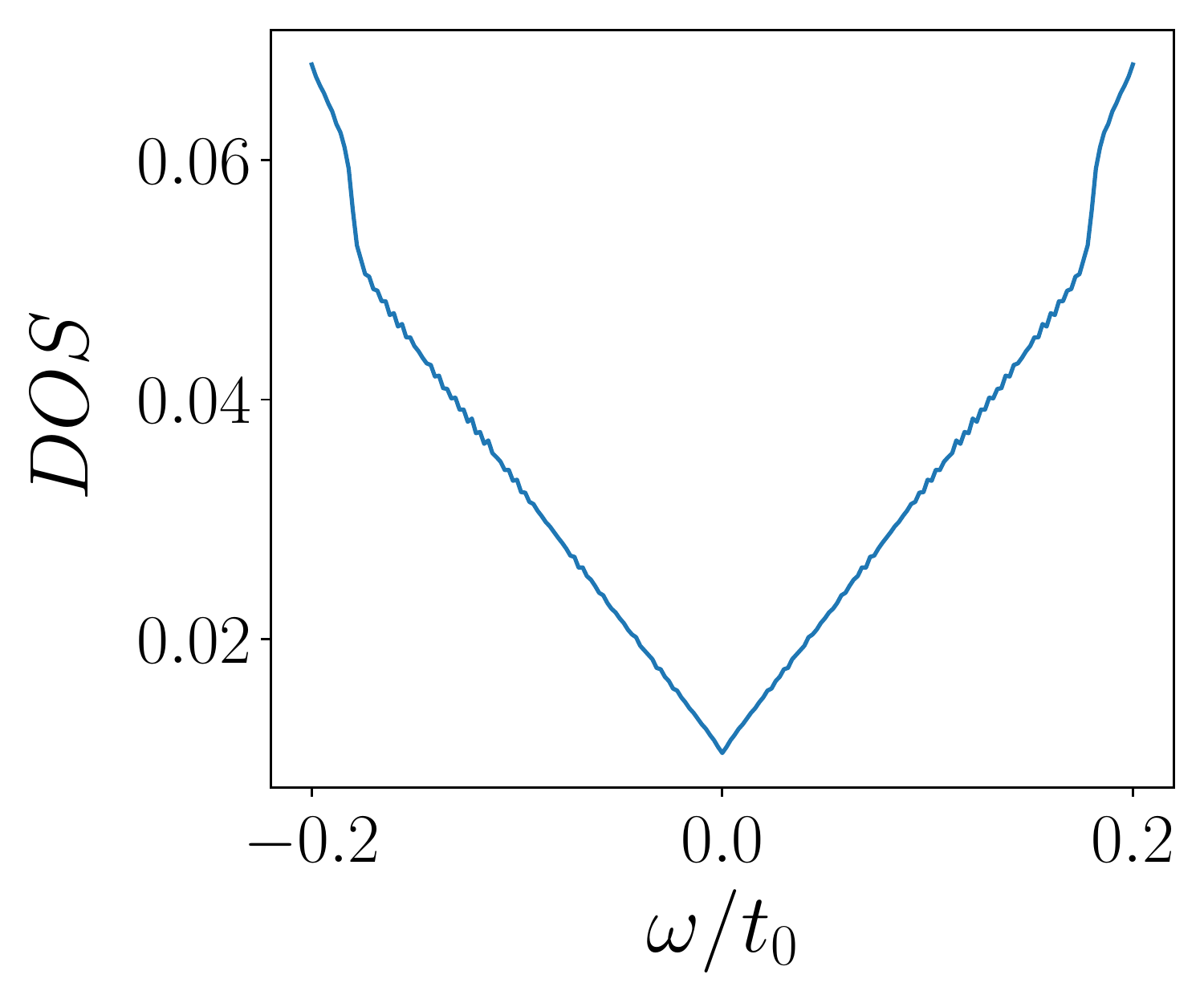}}
\subfloat{\includegraphics[width=0.49\linewidth, trim=0 0 0 0, clip]{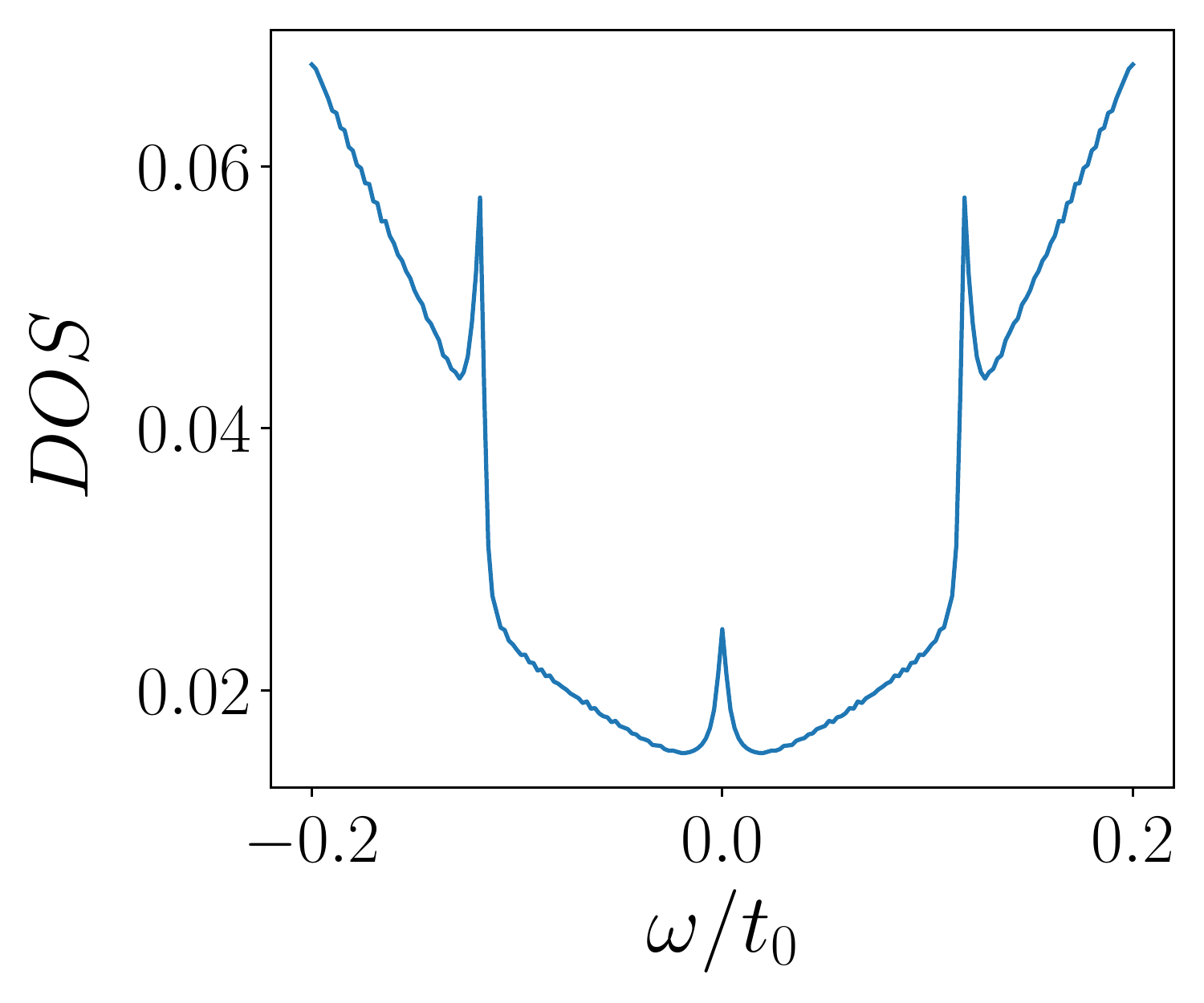}}
\caption{Density of states for ABA (left) and ABC (right) trilayer graphene. For the ABC case, we observe a zero-energy peak due to the presence of the incipient flat band.}
\label{fig:bulkDOS}
\end{figure}

In Fig.~\ref{fig:bulkDOS} we plot the DOS for the ABA and ABC trilayer graphene. Note that the DOS for the ABA graphene shows some quasi-V feature close to zero-energy due to the presence of the combination of the quadratic and the linear band, while the ABC graphene shows a small peak at zero-energy due to the incipient flat band, as well as cusps at energies corresponding to the end of the two quadratic bands.

\section{Ribbons with zigzag edges and interfaces}

\begin{figure}[h]
\centering
\subfloat{\includegraphics[width=0.49\linewidth, trim=0 0 0 20, clip]{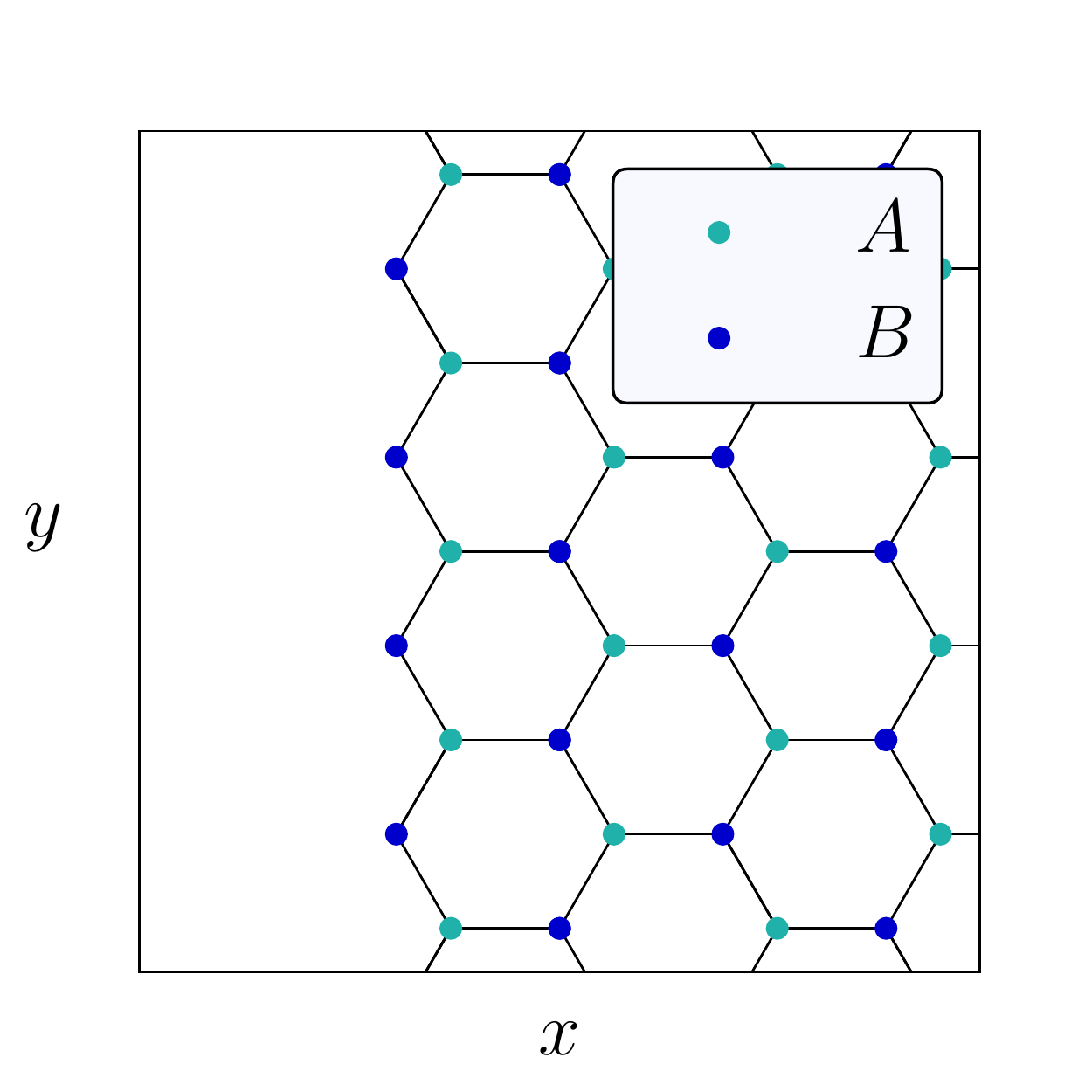}}
\subfloat{\includegraphics[width=0.49\linewidth, trim=0 0 0 20, clip]{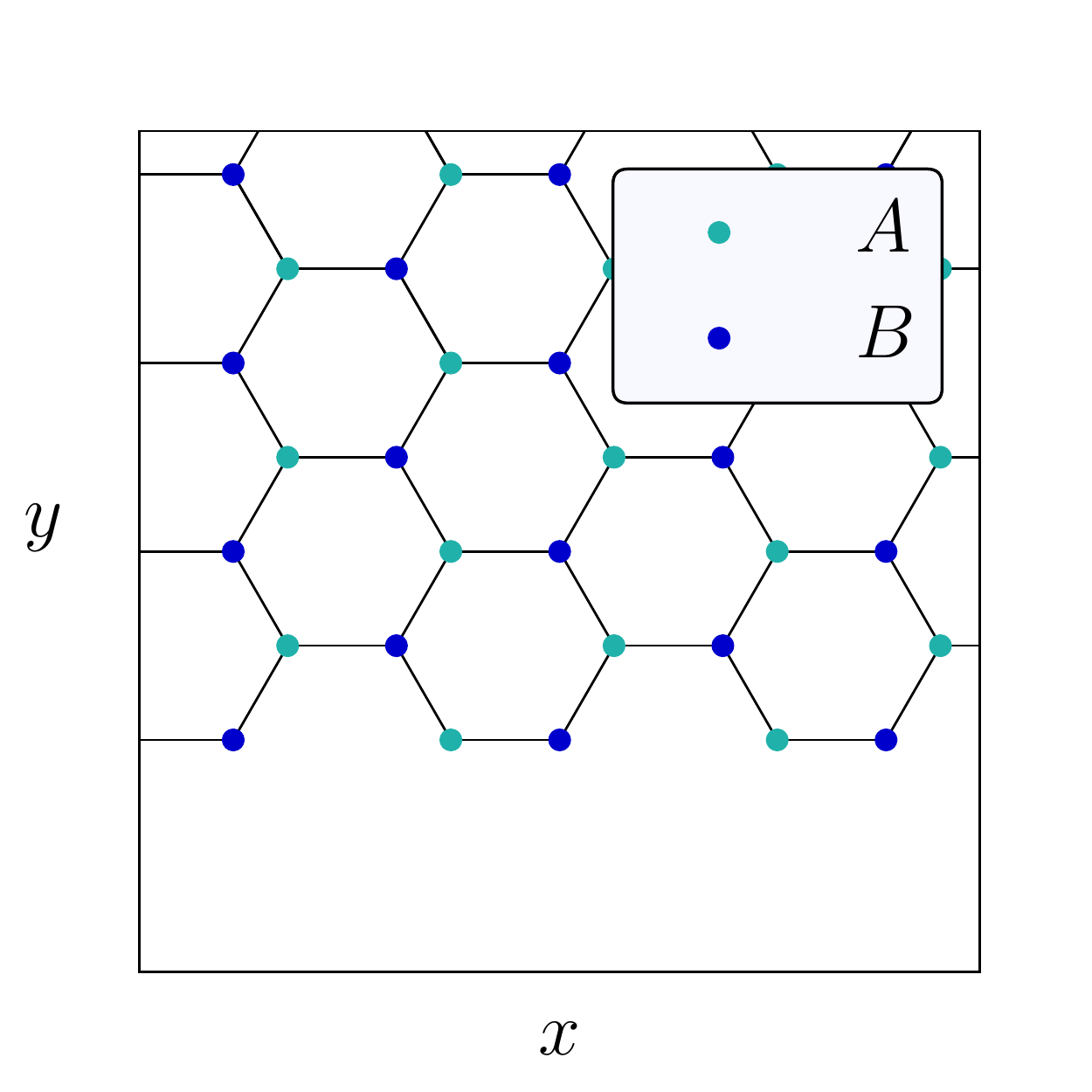}}
\caption{A zigzag (left) and an armchair (right) edge.}
\label{fig:edgesSchematic}
\end{figure}

\begin{figure}[h]
    \centering
    \includegraphics[width = 8.4cm]{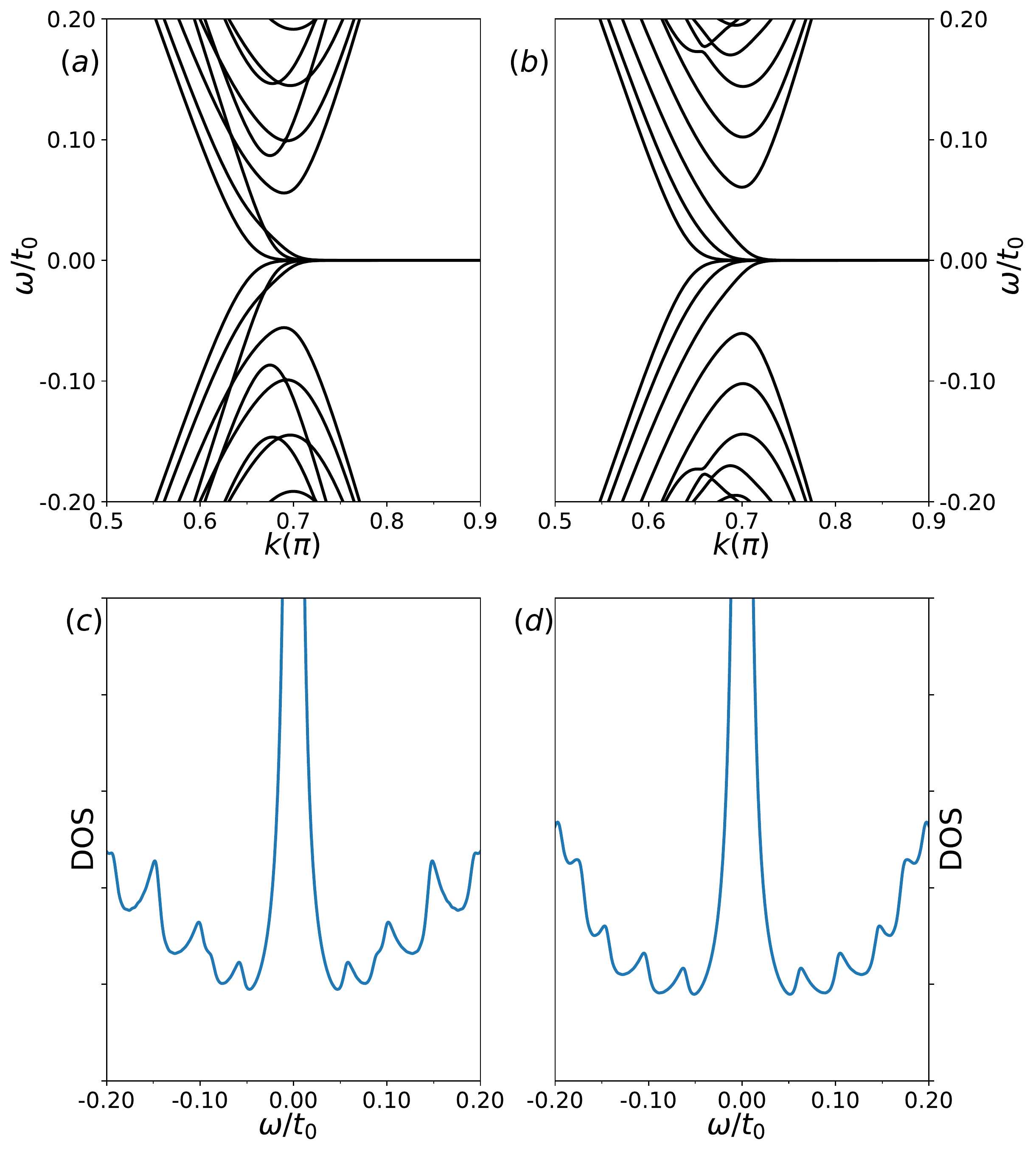}
    \caption{\textbf{(a)} and \textbf{(c)}: Respectively, band structure and density of states for ABA ribbons with zigzag edges. \textbf{(b)} and \textbf{(d)}: Respectively, band structure and density of states for ABC ribbons with zigzag edges. We use ribbons with a width of $78\ a_0$ and we consider a numerical broadening $\eta=2.10^{-3}\ t_0$ for the density of states. In both cases, we see the presence of flat bands at $k=\pi$ corresponding to the localized edge modes which give rise to a pronounced peak in the DOS at $\omega=0$.}
    \label{fig:1}
\end{figure} 

We now consider a ribbon that is finite in one direction ($x$) and infinite in the other ($y$), with zigzag edges (see Fig.~\ref{fig:edgesSchematic}). Along the finite-size direction, the momentum is no longer a good quantum number and the continuous ($k_x$, $k_y$) band structure is replaced by a quantized set of bands. Thus in Fig.~\ref{fig:1} we plot the band structure of zigzag ABA and ABC ribbons. For the ABA structure, we see two types of quantized modes emerging from the linear and the quadratic band, while for the ABC one we can only see the quantized modes emerging from the lowest band.

It is also known that graphene ribbons with zigzag edges exhibit localized edge states. In a trilayer system, these states exist regardless of the stacking order as can be seen in Fig.~\ref{fig:1}: for both ABA and ABC stacking we see that the band structure contains flat bands at zero energy. For each ribbon there are six of those bands, corresponding to the three layers and the two edges of each layer. These states are not topological and can for example move away from zero energy when we apply an electric field.

The corresponding LDOS depicted in the lower panels of Fig.~\ref{fig:1} show clearly the zero-energy edge-state peak, as well as the quantized structure of the band structure. Given the effective 1D character of the ribbons, each band will give rise to a peak in the LDOS at the energy corresponding to the bottom-of-the-band (Van Hove singularity) inflexion. This will yield a series of peaks in the LDOS, as visible in the bottom panel of Fig.~\ref{fig:1}. The overall trend for the LDOS of the ABA/ABC ribbon is consistent with the LDOS for the infinite ABA/ABC trilayer depicted in Fig.~\ref{fig:bulkDOS}, i.e. a linear dispersion for the ABA ribbon, and a linear dispersion with a smaller slope for the ABC ribbon, jumping abruptly at $\omega\approx 0.15t_0$.

\begin{figure}[h]
    \centering
    \includegraphics[width = 8.4cm]{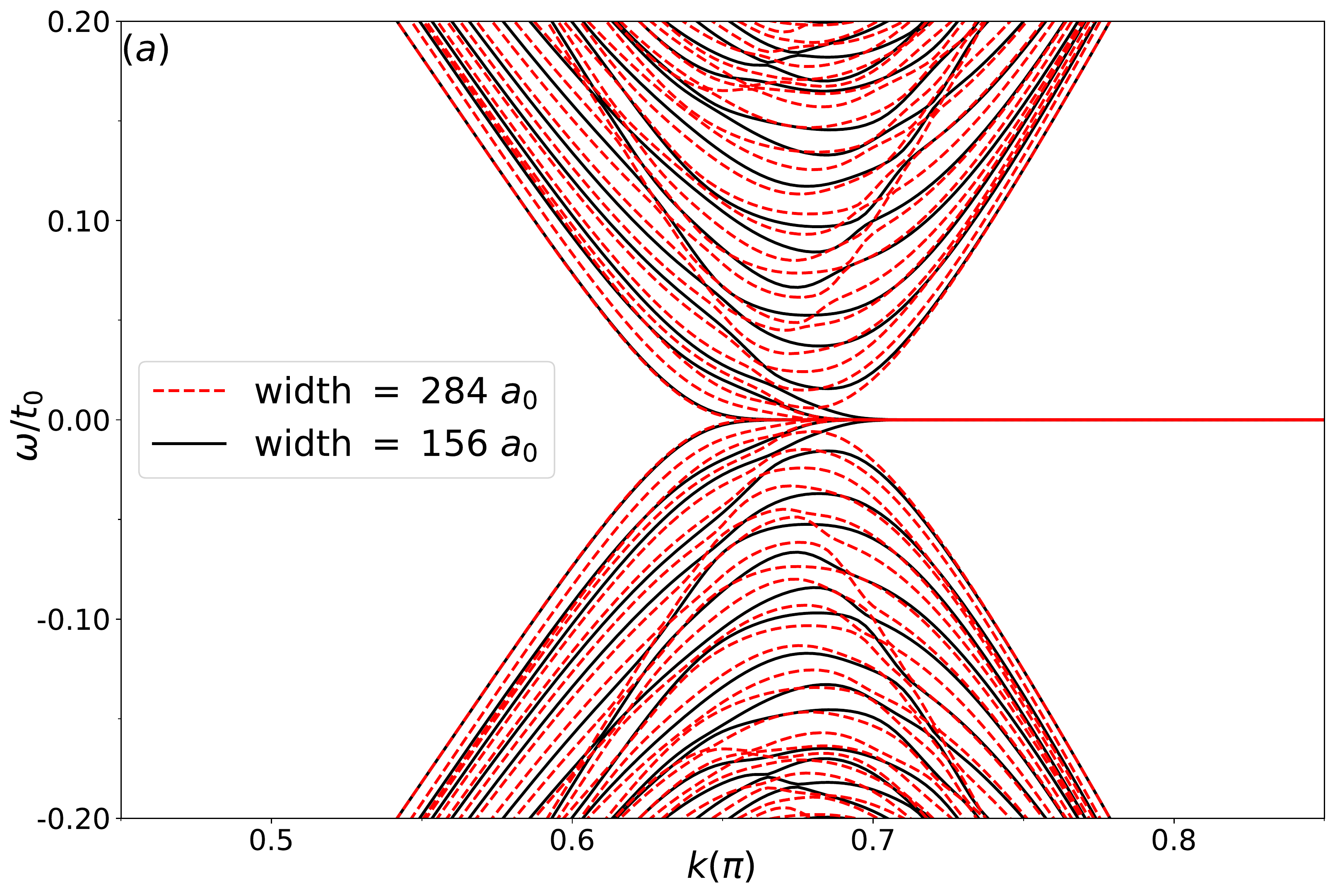}
    \caption{The band structure for a ribbon with a stacking fault, plotted for two different ribbon widths $284 a_0$ (dashed red) and $156 a_0$ (full black). The flat bands and the lowest-energy bands are independent of the width, while the bulk bands strongly depend on it.}
    \label{fig:2}
\end{figure}

When we introduce a stacking fault (see Fig.~\ref{fig:fault}(a)), the band structure is perturbed, as can be seen in Fig.~\ref{fig:2}. Note that because the defect is only localized in the top-most layer, there might be significant hybridization between the two different stacking regions through the bottom layers. Indeed, the bulk bands for the mixed system are no longer similar to the separate ABC/ABA bands but are fully modified by the presence of the defect. To separate the bands coming from the ABC and ABA bulks and those of the interface region we study the effect of the size of the ribbon on the band structure. The idea is that if we change the number of atoms in the bulk, features associated with the edges, as well as with the defect, should remain unchanged. Looking at Fig.~\ref{fig:2}, we see that this is indeed the case for the flat bands around $k=\pi$. These flat bands can easily be identified with states localized on the edges of the ribbon. Moreover, we note that the lowest-energy bands exhibit a dispersion which is independent of the system size even away from the Dirac point.  This size-independent band is peculiar and we argue that it is localized at the interface between the ABA and ABC regions.

\begin{figure}[h]
    \centering
    \includegraphics[width = 8.4cm]{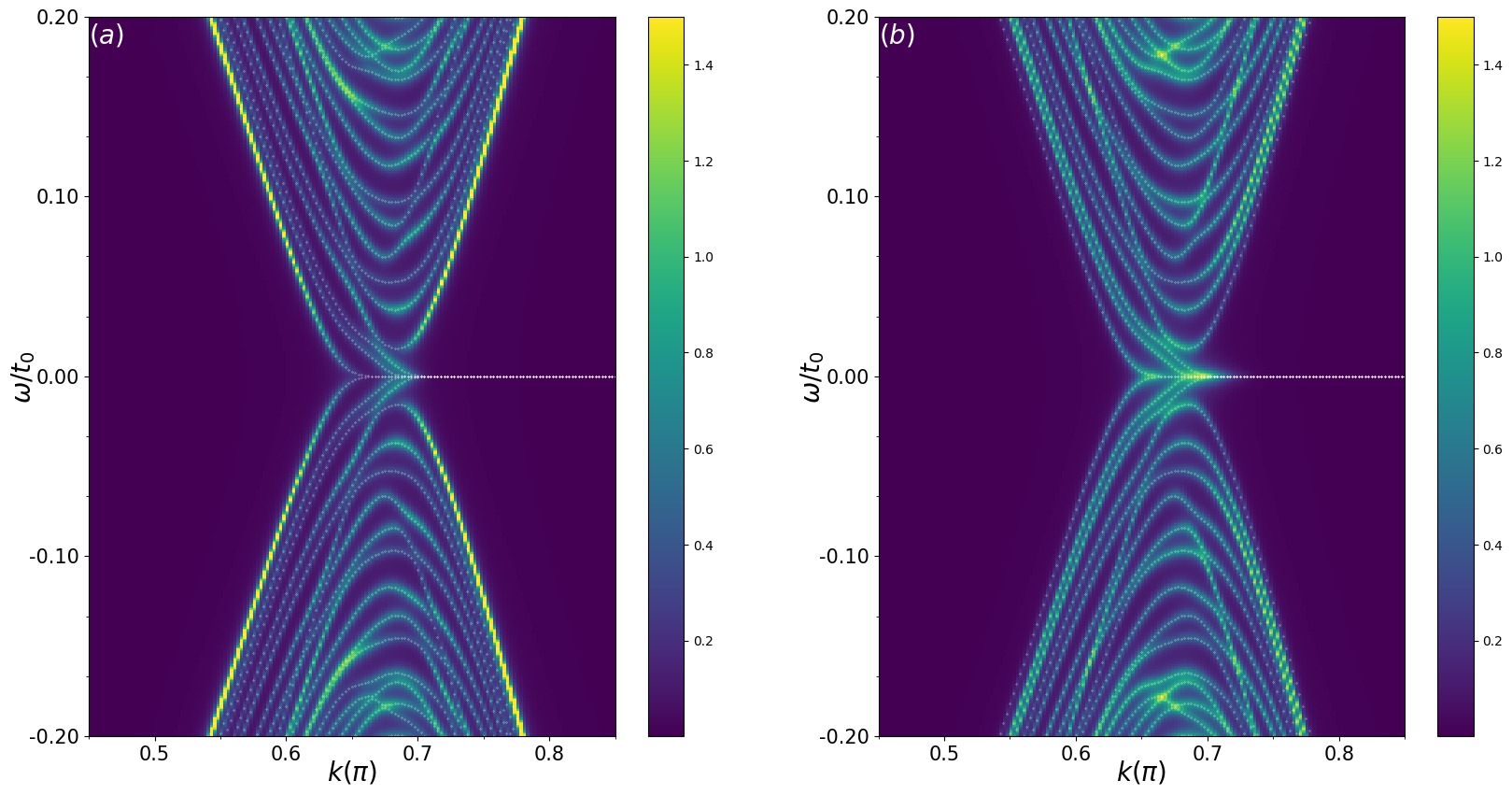}
    \caption{\textbf{(a)} The eigenstate spectral weight evaluated on sites in the stacking fault. There is a significant weight in the two lowest positive- and negative-energy bands, indicating that these states are localized on the defect. \textbf{(b)} Averaged spectral weight evaluated on sites in the bulk (average of ABA and ABC sides). The ribbon's width is $156\ a_0$ and we used a numerical broadening $\eta = 3.10^{-3}\ t_0$}
    \label{fig:3}
\end{figure}

To confirm this observation we study the spectral weight of the different bands as a function of the position on the ribbon. Fig.~\ref{fig:3}(a) shows the spectral weight averaged over two unit cells in the stacking fault, while Fig.~\ref{fig:3}(b) shows the averaged spectral weight for sites in the bulk on both the ABA and ABC sides of the ribbon. We see that the interface spectral weight is significantly larger for the lowest dispersing energy band, at both positive and negative energies. In contrast, the bulk spectral weight is rather equally distributed on the quantized multiple bands in the Dirac cone with a marked absence of weight on the flat band (localized at the exterior edges) and on the lowest energy band (localized at the interface). This provides an extra indication that the bands we identified as size-independent are localized on the defect. It is also interesting to note that the situation is less clear for momenta close to the Dirac point where the spectral weight is once again evenly distributed (Fig.~\ref{fig:3}(a)) and there are no signs of localized states.

\begin{figure}[h]
    \centering
    \includegraphics[width = 8.4cm]{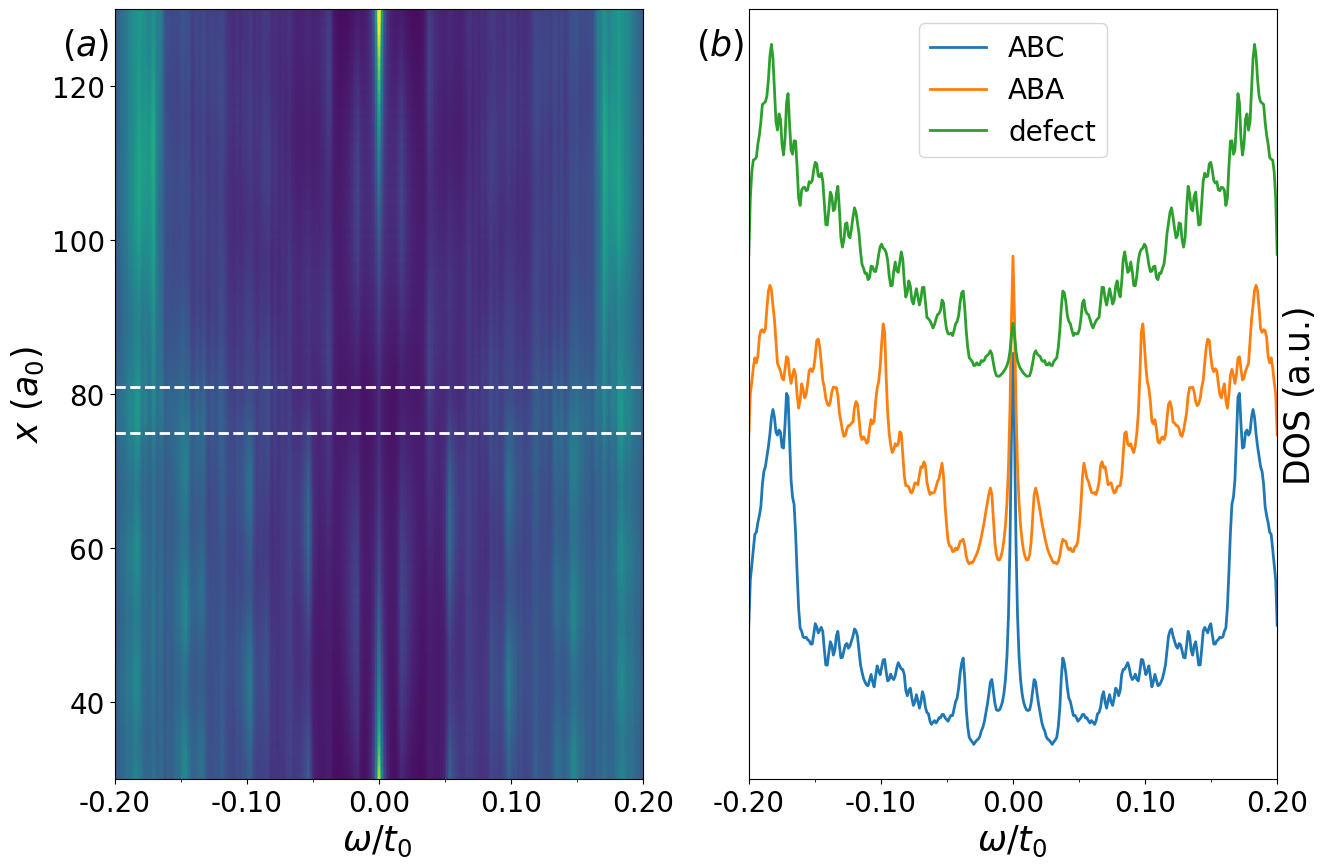}
    \caption{\textbf{(a)} Local density of states as a function of energy and position along the ribbon. The ABC region corresponds to $x>80$ (top of the picture), while the ABA region to $x<75$ (bottom of the picture). We left out the ribbon's edges where there is a high density of states due to the edge states in the zigzag geometry. The white dashed lines indicate the position of the stacking fault. \textbf{(b)} Density of states at specific sites. The DOS on the ABA ($x=39$) and ABC ($x=119$) sides are taken far from the defect ($x=78$), in the bulk. The ribbon's width is $156\ a_0$ for both figures. DOS lines are shifted for readability. We used a numerical broadening $\eta = 2.10^{-3}\ t_0$.}
    \label{fig:4}
\end{figure}

Lastly, we look for possible signatures of these states in the density of states as a function of position along the ribbon. We see in Fig.~\ref{fig:4}(a) that the local density of states (LDOS) at zero energy is dominated by the edge states. This is due to the flat nature of the edge bands give which rise to a high density of states close to $\omega = 0$. This is localized both on the edge but also exhibits significant leaks in the bulk, for the not-too-large system size that we consider. To get a better understanding in Fig.~\ref{fig:4}(a) we plot the LDOS as a function of energy in the stacking fault, as well as in the ABA and ABC regions. We thus note that the LDOS in the defect (the region between the white dashed lines in Fig.~\ref{fig:4}(a)), as well as in the ABA and ABC regions far from the edges, does still present a significant peak at $\omega = 0$ that is due to the zigzag edge modes. This makes it hard to separate for example the effect of the bulk ABC zero-energy flat band from that of the zigzag edges.

The dependence of the DOS in the defect on energy (green curve in Fig.~\ref{fig:4}(b)) exhibits a roughly linear dependence on energy, similar to that of the bulk ABA region depicted in Fig.~\ref{fig:bulkDOS}; this is consistent with the fact that de defect spectral weight is maximum on a pair of bands that disperse linearly away from $k=0$ (Fig.~\ref{fig:3}(a)). These bands decay in intensity close to zero energy, consistent with the reduction in the LDOS of the defect at small energies, as shown in Fig.~\ref{fig:4}(b). The bulk ABA preserves also a linear dependence (orange curve in Fig.~\ref{fig:4}(b)), modulated by the finite-size effects. The ABC bulk LDOS (Fig.~\ref{fig:4}(b) blue curve) also preserves the average ABC uniform bulk dependence depicted in Fig.~\ref{fig:bulkDOS}, i.e a linear dependence with a lower slope for $0<\omega<0.15$ before increasing rapidly. 


Note that the LDOS on the interface has roughly the same intensity as in the bulk, the interface does not seem to generate significant interface states localized at the position of the defect.

\section{Ribbons with armchair edges and interfaces}

\begin{figure}[h]
    \centering
    \includegraphics[width = 8.4cm]{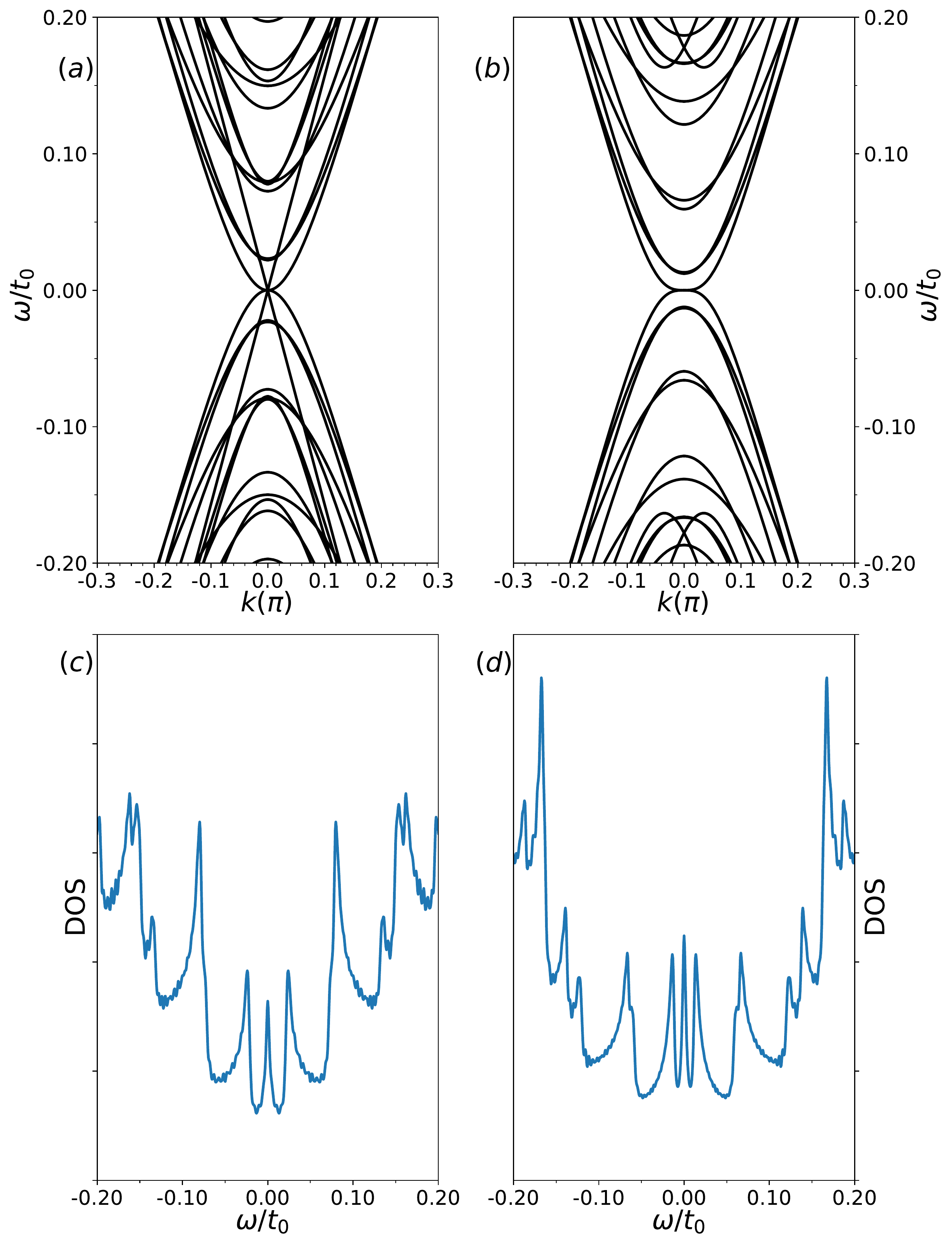}
    \caption{\textbf{(a)} and \textbf{(c)} Band structure and respectively density of states for ABA ribbons with armchair edges. \textbf{(b)} and \textbf{(d)} Band structure and respectively density of states for ABC ribbons with armchair edges. We used ribbons with a width of $59\ a_0$, and a numerical broadening $\eta=2.10^{-3}\ t_0$ for the density of states. Both stacking seem to exhibit a zero-energy peak in the LDOS, but the ABC stacking one is more robust due to the flatter dispersion around $k=0$.}
    \label{fig:5}
\end{figure}

We now perform the same analysis on ribbons with armchair edges. The geometry of the stacking fault is depicted in Fig.~\ref{fig:fault}(b). In the case of armchair ribbons there are no edge states, and the band structures for the two different stackings are shown in Fig.~\ref{fig:5}(a) and (b). Compared to the zigzag case, there is a stronger contrast between the two band structures: for the ABA stacking the linearly-dispersing and the lowest quadratically-dispersing bands seem to be preserved even in the presence of the quantization. The band structure of the ABC-stacked ribbon seems to be affected also much less by the quantization, preserving the $k^3$-dispersing bands touching at $\omega =0$; we thus expect some reminiscence of the flat band signatures (such as a zero-energy peak in the DOS) in this type of geometry. Indeed, while both stackings give rise to zero-energy peaks due to the zero-energy van Hove singularities corresponding to the bottom-of-the-band inflexions in the now effectively-1D energy bands, the one corresponding to the ABC stacking is more robust.

\begin{figure}[h]
    \centering
    \includegraphics[width = 8.4cm]{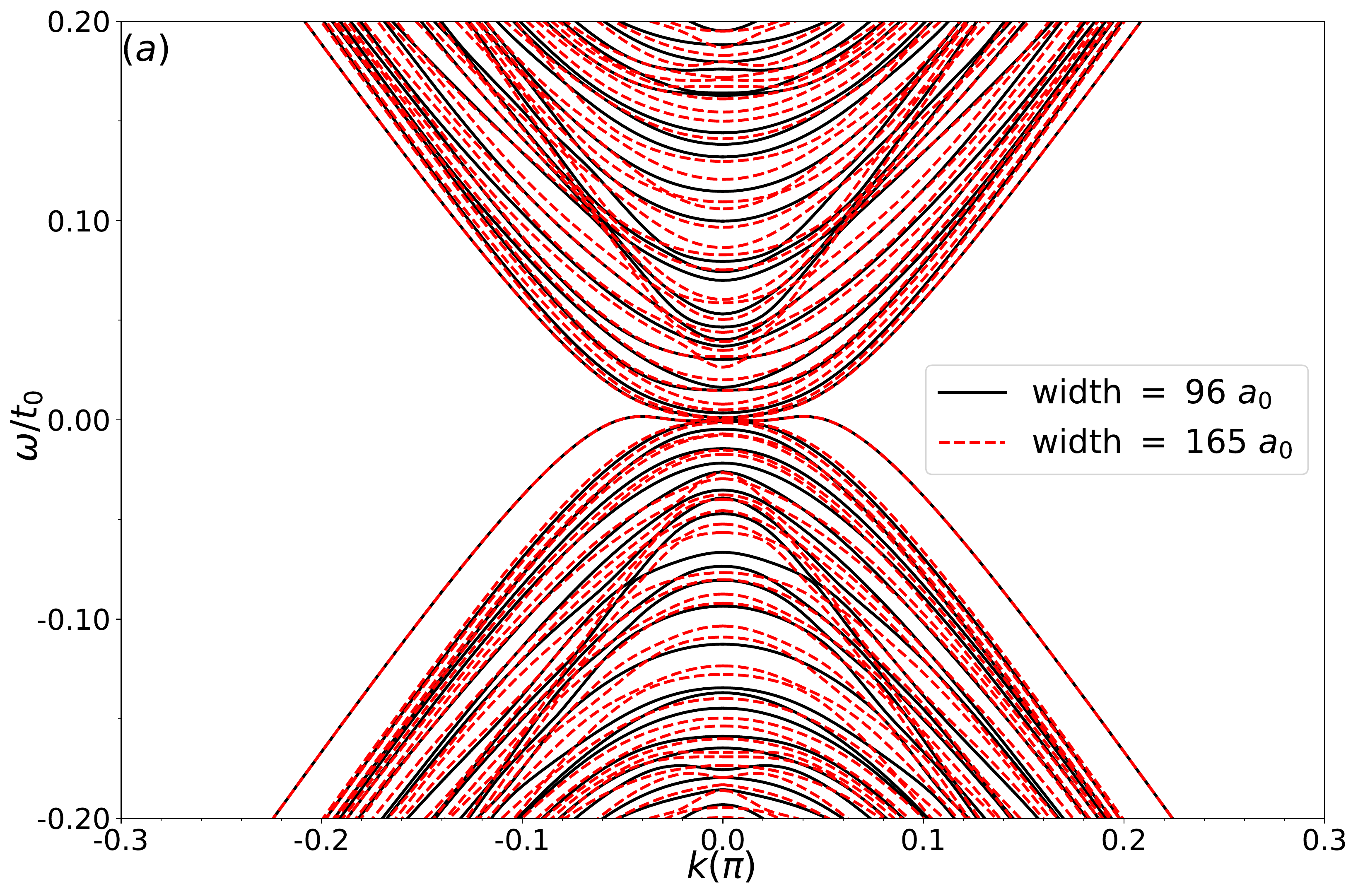}
    \caption{ The band structure for an armchair ribbon with a stacking fault for two different ribbon widths. The lowest-energy dispersing bands are the same for the two different widths, while the bulk bands are different.}
    \label{fig:6}
\end{figure}

The band structure of a ribbon with a stacking fault and armchair edges is shown in Fig.~\ref{fig:6} for two different ribbon widths. Once again the band structure of the combined system cannot be directly traced back to the individual band structures for the ABA and ABC ribbons. Moreover, we can clearly see that the combined-system band structure is not symmetric with respect to $\omega=0$. This symmetry is usually a consequence of the sublattice symmetry that exists in the honeycomb lattice and is broken here due to the particular shape of the impurity (see Fig.~\ref{fig:fault}(b)). The stacking fault gives rise to a distinct signature in the band structure: we can see the formation of two dispersing bands (one for positive and one for negative energies) which seem to be unaffected by the size of the system. This feature indicates that these may indeed be interface bands, and not bulk bands, same as in the zigzag ribbon case. These bands are once more the lowest in energy in the spectrum and have a relatively flat dispersion near $k=0$. 

\begin{figure}[h]
    \centering
    \includegraphics[width = 8.4cm]{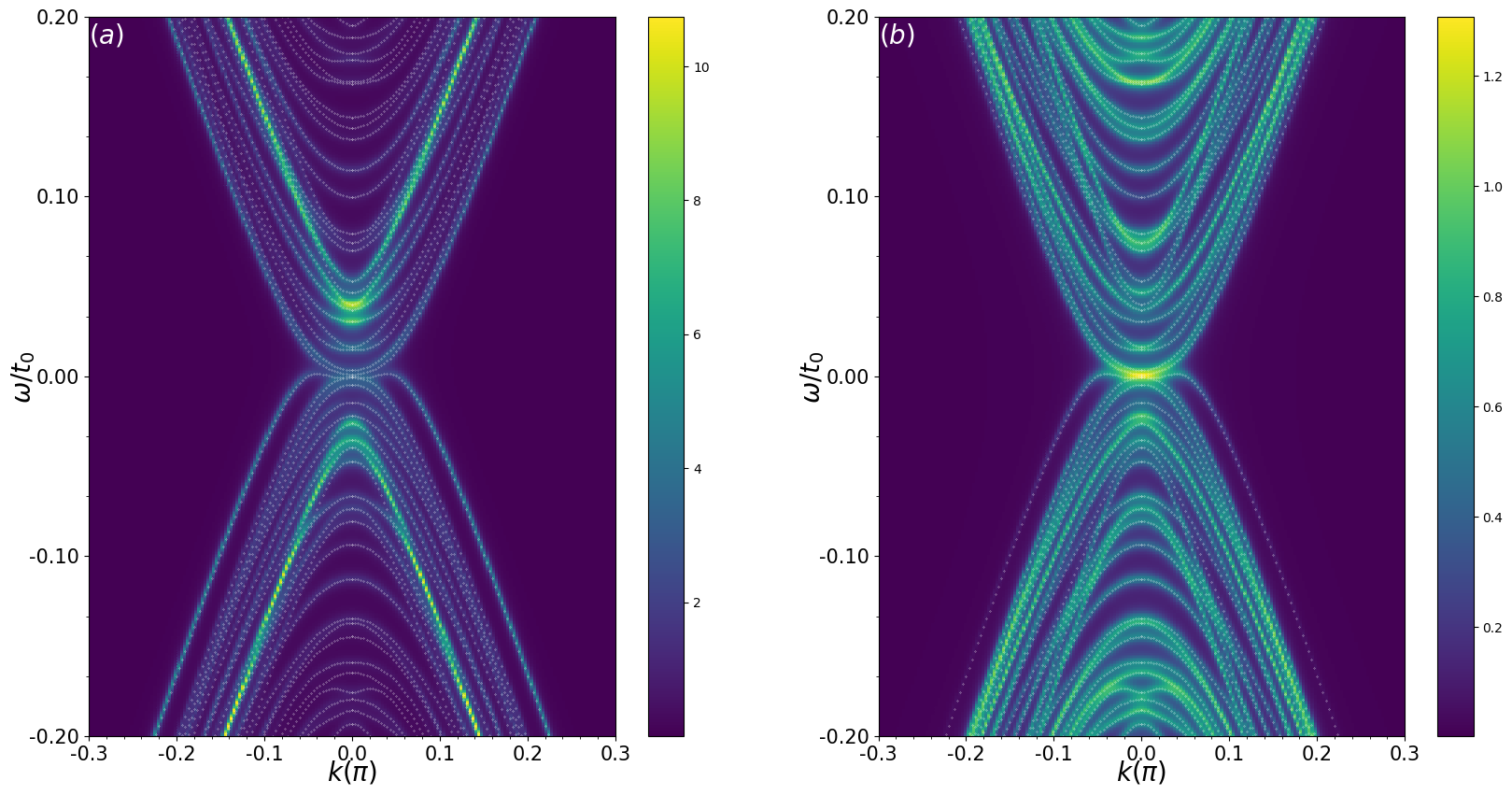}
    \caption{\textbf{(a)} The eigenstate spectral weight evaluated on sites in the stacking fault. There is a significant weight in the two lowest positive- and negative-energy bands, indicating that these states are localized on the defect. Moreover, reminiscent segments from the linear band arising for a clean ABC structure also shows a very strong intensity on the defect; however this band is now gapped. \textbf{(b)} Averaged spectral weight evaluated on sites in the bulk (average of ABA and ABC sides). Note the large difference between the scales of the two figures indicating a strong intensity for the states localized on the defect. The ribbon's width is $96\ a_0$ and we used a numerical broadening $\eta = 3.10^{-3}\ t_0$}
    \label{fig:7}
\end{figure}

Same as in the zigzag case we also calculate the spectral weight on sites located in the stacking fault, and we compare it to the spectral weight in the bulk. This is shown in Fig.~\ref{fig:7}: we can see that the interface spectral weight is localized most strongly on the two lowest positive- and negative-energy bands (Fig.~\ref{fig:7}a). Moreover, segments from the central band, reminiscent of the linear ABA band, exhibit a very strong intensity at the ABA/ABC interface. On the other hand, the spectral weight in the bulk (Fig.~\ref{fig:7}b) seems evenly distributed, except on the lowest energy band which has very little weight, confirming that this band is an interface band and not a bulk band. We note the strong difference in the scale of the spectral weight between the two figures, indicating that the interface states have much more intensity than the bulk states.

\begin{figure}[h]
    \centering
    \includegraphics[width = 8.4cm]{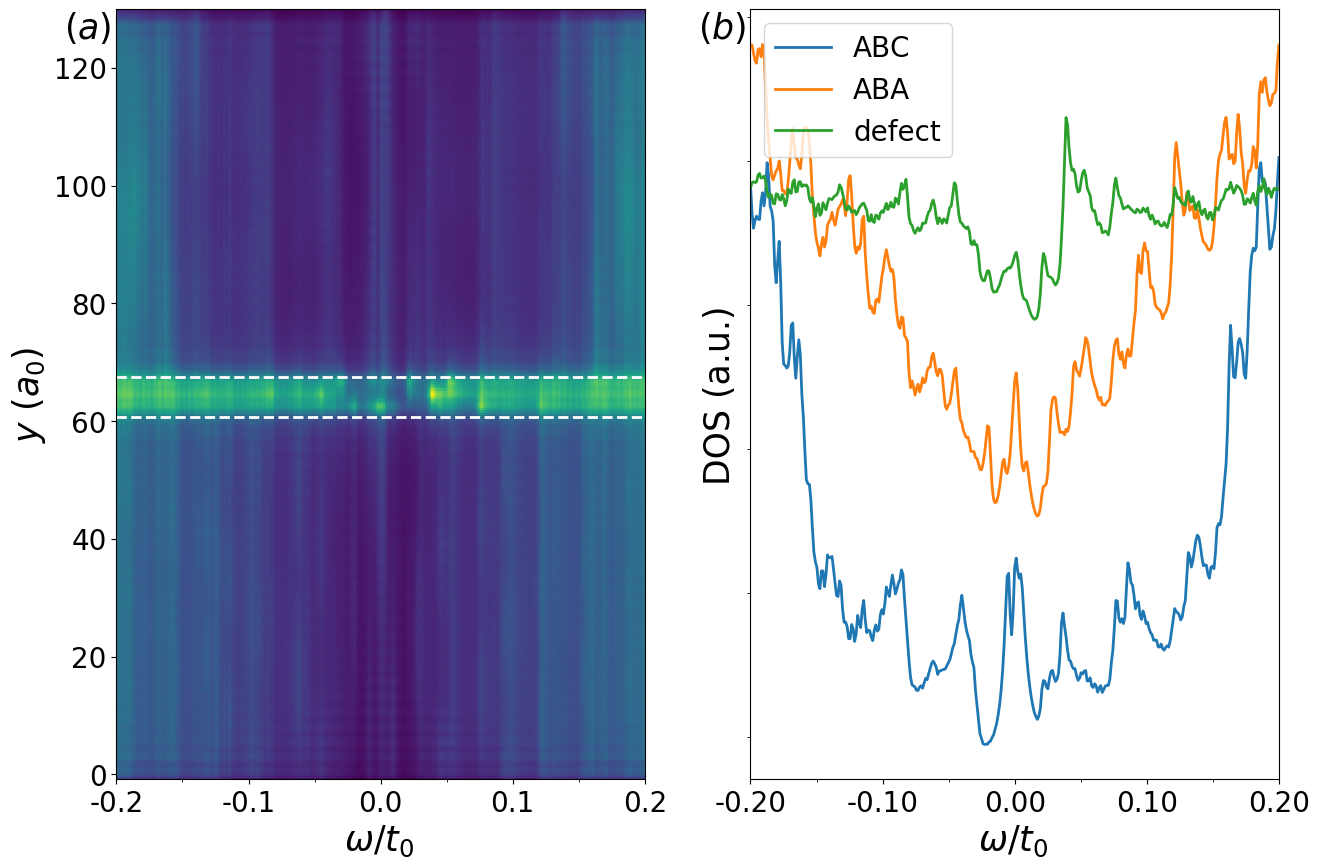}
    \caption{\textbf{(a)} Local density of states as a function of energy and position along the ribbon. The ABC region corresponds to $y>70$ (top of the picture), while the ABA region to $y<60$ (bottom of the picture). The white dashed lines indicate the position of the defect. \textbf{(b)} Density of states at specific sites. The DOS on the ABA ($y=31$) and ABC ($y=98$) sides are taken far from the defect, in the bulk. The DOS taken on the defect ($y=65$) is divided by 10 for scaling reasons. The ribbon's width is $96\ a_0$ for both figures. DOS lines are shifted for readability. We used a numerical broadening $\eta = 2.10^{-3}\ t_0$. }
    \label{fig:8}
\end{figure}

We also perform an analysis of the LDOS as a function of position (Fig.~\ref{fig:8}(a)). This turns out to be very different from the case of the zigzag ribbon, as there are no zero-energy edge states and also there is a higher spectral weight for the states localized on the impurity. We note first that the  LDOS in the ABA and the ABC bulk becomes also asymmetric between the positive and negative energies. The ABA LDOS (Fig.~\ref{fig:8}(b) orange line) shows only small modifications from the previously studied situations (i.e a linear background plus quantized peaks, see Fig.~\ref{fig:5}). The ABC bulk (Fig.~\ref{fig:8}(b) blue line) also follows the structure in Fig.~\ref{fig:5}, exhibiting a a linear slope for $0<\omega<0.15$ before increasing rapidly, as well as a strong zero-energy peak corresponding to the flat band, but with a more pronounced asymmetry between the positive and negative energies, and a reduction in the LDOS at small negative energies.

 The averaged DOS in the defect (Fig.~\ref{fig:8}(b) green line) shows a reduction in the LDOS around $\omega=0$, and sharp peaks for positive energies $\omega\approx 0.05$. This is in agreement with the momentum space analysis (Fig.~\ref{fig:7}(a)) where we can see that the spectral weight is maximum on a pair of gapped bands which are reminiscent of the linear ABA bands: the sharp peaks arise at the gap edge of these bands. We also see that the defect DOS is not symmetric with respect to $\omega = 0$ as expected from the band structure in Fig.~\ref{fig:6}.  Also we note the large intensity of the LDOS in the defect which persists at all energies.

\section{Finite-size configuration}
In this section we consider instead of an infinite ribbon a finite-size ABA-ABC ribbon. Once again we look at systems with zigzag or armchair terminations, but with a finite size in both the $x$ and $y$ directions. We thus expect a quantization of the continuous interface bands present in the infinite ribbon configurations, and the appearance of quantized interface levels. Since there is no longer any good momentum quantum number there is no band structure to analyze, but only the LDOS as a function of energy and position.

\begin{figure}
    \centering
    \includegraphics[width = 8.4cm]{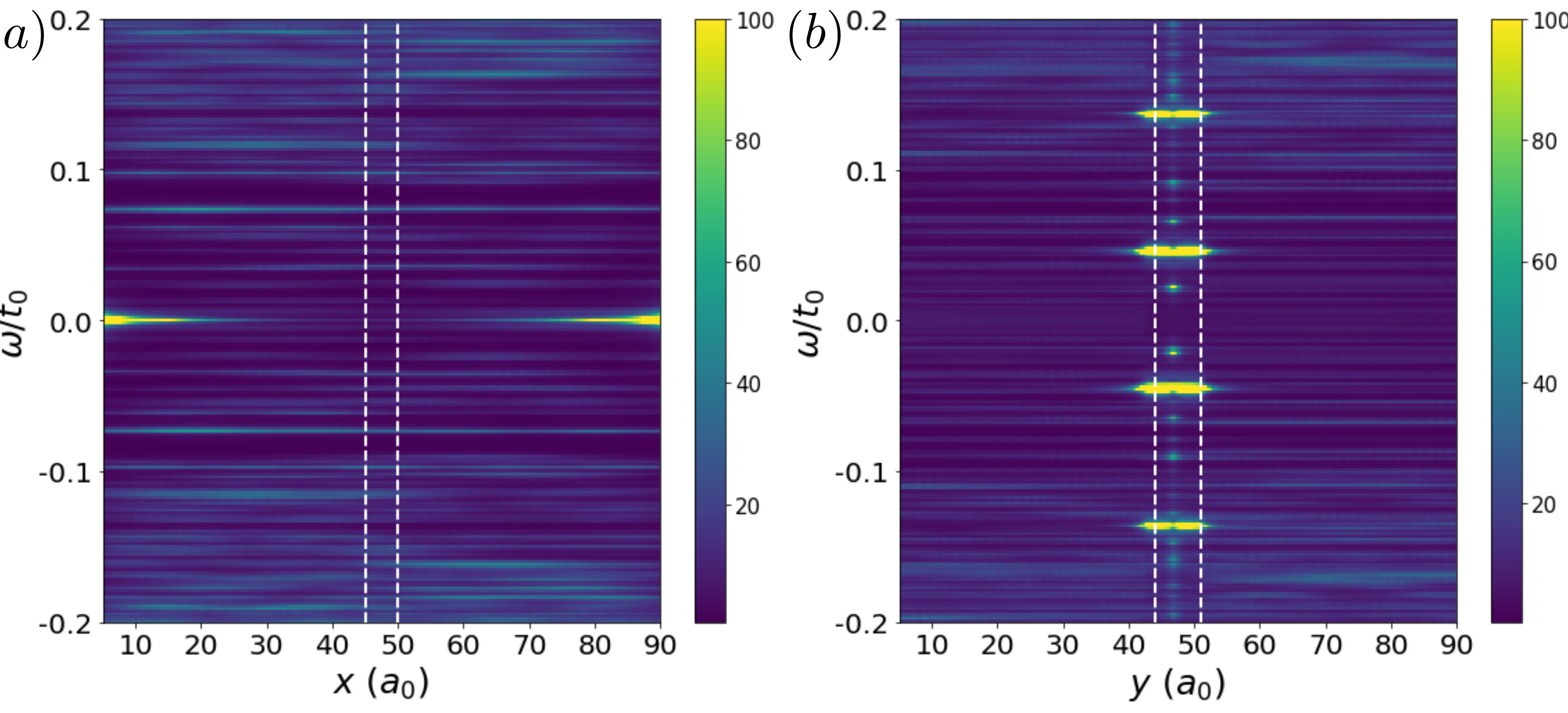}
    \caption{LDOS as a function of energy and position for a finite-size system with zigzag edges (a) and armchair edges (b). The white dashed lines indicate the position of the defect. The ABA region corresponds to $x<45$ (left of the stacking fault), while the ABC region to $x>50$ (right of the stacking fault). We used a numerical broadening $\eta = 1.10^{-3}\ t_0$, and for the zigzag ribbon we have excluded the leftmost and rightmost edges. The length of the ribbons (perpendicular to the impurity) is $95a_0$, while the width is $35a_0$ in both geometries.}
    \label{fig:9}
\end{figure}

We start by studying the case of zigzag edges. In Fig.~\ref{fig:9}(a) we plot the LDOS as a function of energy along a cut perpendicular to the direction of the stacking fault. We observe some predominant features. First we note that the edge states, localized at the exterior edges of the sample, continue to exhibit a strong spectral weight even in the finite configuration, this is expected from previous studies of graphene. We can see here that these states also penetrate quite significantly in the bulk ($\sim 30a_0$). The second relevant feature is the formation of the quantized states localized at the ABA/ABC interface, these are just a result of the quantization of the interface bands observed in the previous section. Their intensity once more is not more significant than that of the bulk, indicating that the zigzag interface does not generate strong localized interface states.


We now turn to the case of armchair edges, and in Fig.~\ref{fig:9}(d) we plot the LDOS as a function of energy and position along the $y$ axis. Here we do not have any zero-energy modes as the armchair edges do not exhibit any localized states. We can see however very clearly the quantization of the energy levels in the stacking fault. The spectral weight associated with these localized states is higher than the spectral weight for delocalized states in the ABC or ABA bulk. This is in agreement with the results for the semi-infinite ribbon presented in Section IV. 
The limited size for the bulk ABA and ABC stacking regions does not allow us to have a good quantitative understanding for the low-energy DOS outside the impurity.

In the Appendix we study a different finite-size configuration, that of  a square system for which the outer region has an ABC stacking while the inner region an ABA stacking, which confirms the quantization of the edge modes observed here, as well as a significant dependence of the intensity of the edge modes with the details of the stacking fault configuration, for both zigzag and armchair interfaces.

\section{Conclusion}
We have studied the formation of interface states between ABC and ABA trilayer graphene regions. We have considered both zigzag and armchair junctions and both infinite ribbons and finite-size configurations. 
We noticed the formation of interface dispersing bands, which differ between the zigzag and armchair cases. They are very intense for the armchair geometry, while in the zigzag case they do not show a significant intensity with respect to the bulk bands. The local density of states also clearly indicates that these states are localized in the stacking faults. The corresponding bands transform into a set of discrete quantized levels for a fully finite-size configuration. We believe that these states are non-topological, given their strong dependence on the type of interface (zigzag or armchair), as well as on the details of the stalking fault interface. This is consistent with the fact that there is no notion of a non-zero topological invariant in graphene in the absence of an electric field. Our observations should be important for experiments concerning interfaces between ABA and ABC graphene nanoribbons, as well as graphene regions of a particular stacking embedded in a bulk with a different stacking.

\acknowledgments

C.P. and M.G. would like to acknowledge financial support from the ERC, under grant agreement AdG-694651-CHAMPAGNE. We would like to thank Claude Chapelier and Vincent Renard for bringing up the topic of ABC/ABA junctions and for fruitful discussions.

\bibliography{biblio}

\newpage
\widetext
\appendix
\section{Square system}
\begin{figure}
    \centering
    \includegraphics[width = 8.4cm]{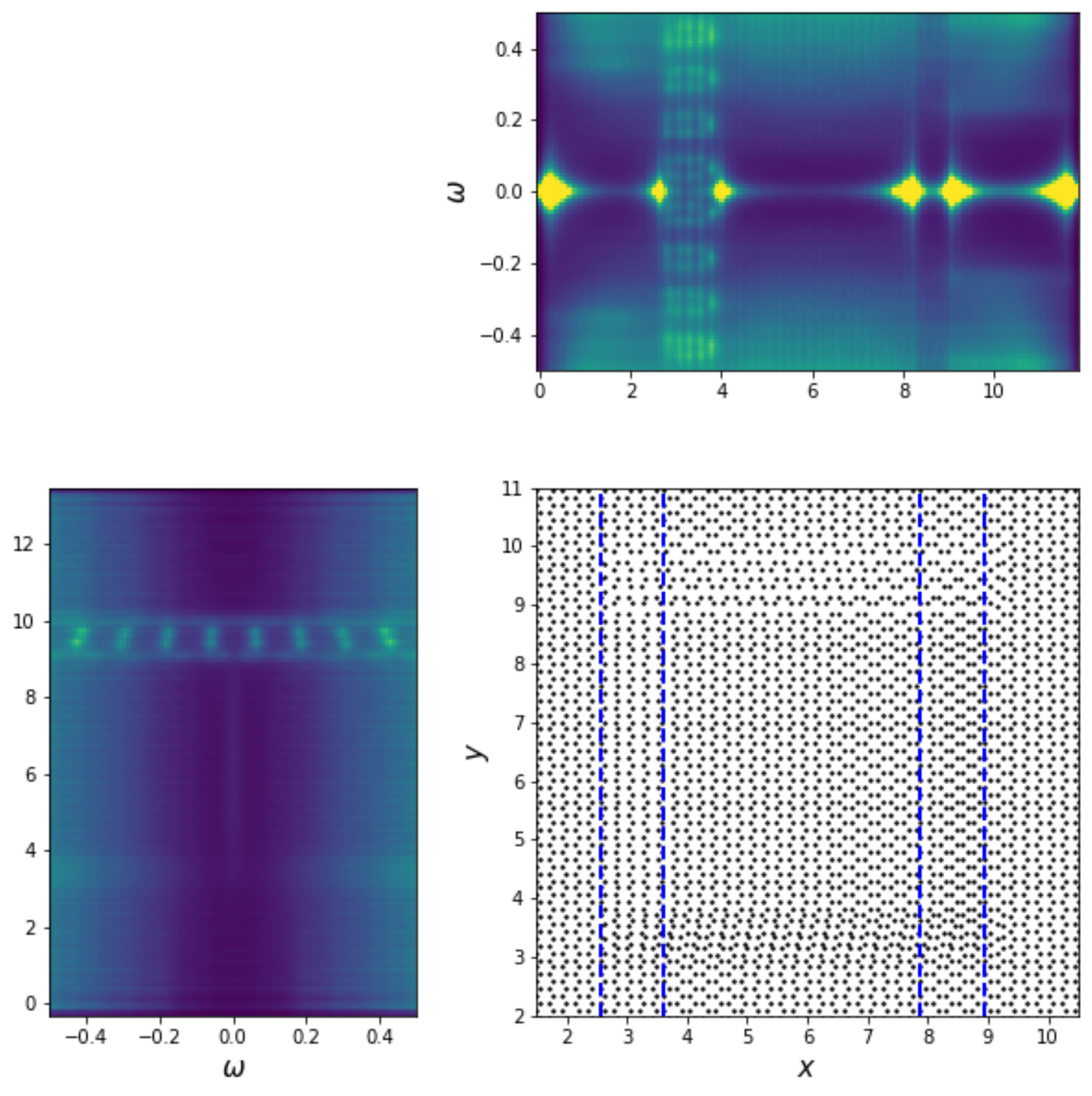}
    \caption{\textbf{bottom right}: Top layer of a graphene trilayer with ABC stacking with an ABA island in the middle. The stacking faults have a zigzag termination along the $x$ axis and an armchair one along the $y$ axis. The blue dashed lines indicate the position to the defects. \textbf{top right} LDOS as a function of energy and position along the $x$ axis, for a fixed value of $y\approx 7$. The stacking faults have a zigzag geometry and correspond respectively to a 'stretching' and a 'compressing' defect for the left-side and right-side impurities. We observe strongly localized quantized states in the stretching case but the DOS in the compressing case is featureless, consistent with the results presented in the main text. \textbf{bottom left}: LDOS as a function of energy and position along the $y$ axis, for a fixed value of $x\approx 6$. The stacking faults have an armchair geometry and correspond to the impurity studied in the main text (top side) and its complementary (bottom side). We observe strongly localized quantized states on the top side, consistent with the results in the main text, while the complementary  defect does not show any strong signature.}
    \label{fig:A1}
\end{figure}

We present in this appendix the result of a tight-binding calculation on a square system for which the outer region has an ABC stacking while the inner region an ABA stacking. The transition between the two types of stacking is made by four defects, each on a side of a square, among which two have an armchair configuration and two have a zigzag configuration. The positions of the atoms in the top layer, that are responsible for the stacking faults, is presented in Fig.~\ref{fig:A1}. The right side of the impurity square has the same atomic configuration as the zigzag stacking fault presented in the text, while the top side corresponds to the armchair one described in Fig.~\ref{fig:fault}. For the other sides we have chosen complementary impurities, such that the left side is a 'stretching' impurity, i.e. the transition between the two stackings is realized not by shortening the distance between the atoms inside the interface region (as we did for the 'compressing' impurity in the main text), but by increasing it. For the bottom side we have taken an impurity for which the atoms are shifted to the left, rather than to the right, oppositely to the armchair interface studied in the main text. These choices insure that we recover the same stacking and an uniform bulk outside the impurity square in all four directions.\\

In Fig.~\ref{fig:A1} top-right and bottom-left panels we plot the LDOS as a function of position and energy, along both the zigzag ($x$) direction and along the armchair ($y$) direction. We can clearly see that even if they are created along the same type of interface (armchair or zigzag), the complementary impurities have significantly different effects on the density of states. Thus for the zigzag 'stretching' defects as well as for the armchair defects presented in the main text (left and top side) we recover strong quantized states localized on the interface. The situation is however reversed in the case of a zigzag 'compressing' defect, as well as for the armchair defect complementary to that of the main text (right and bottom sides) where the LDOS does not show strong localized states on the impurity. This is a strong indication that the states that we found are very sensitive to lattice effects, and thus they are not topological states and can be tuned in and out of existence by modifying the lattice structure, same as zigzag edge states in monolayer graphene.

\end{document}